\documentclass{rmf-d}
\usepackage{nopageno,rmfbib,multicol,times,epsf,amsmath,amssymb,cite}
\usepackage{slashed}
\usepackage[utf8x]{inputenc}
\usepackage[]{caption2}
\usepackage{graphics,graphicx}
\usepackage{graphicx,color}
\usepackage{subfigure} 
\usepackage{wrapfig} 
\usepackage{epstopdf}
\graphicspath{{figuras/}}

\def\rmfcornisa{Research OR Education of Physics \hfill\rmf\ {\bf ??} (*?*) ???--???
\hfill MES? A\~NO?}

\def\rmfcintilla{{\it Rev.\ Mex.\ Fis.\/} {\bf ??} (*?*) (????) ???--???}
\clearpage \rmfcaptionstyle 
\begin{document}
\title{HOLOGRAPHIC CONFINEMENT OF THE SOLITARY QUARK
\vspace{-6pt}}
\author{ Saulo de Mesquita Diles$^a$, Miguel Angel Mart\'in Contreras$^b$ and Alfredo Vega$^b$   }
\address{ $^a$Campus Salin\'opolis, Universidade Federal do Par\'a,
68721-000, Salin\'opolis, Par\'a, Brazil  }
\address{$^b$Instituto de F\'\i sica y Astronom\'ia, Universidad de Valpara\'iso, A. Gran Breta\~na 1111, Valpara\'iso, Chile }

\author{ }
\address{$^*$smdiles@ufpa.br }
\maketitle
\recibido{day month year}{day month year
\vspace{-12pt}}
\begin{abstract}
\vspace{1em}Single quarks moving in the vacuum of confining gauge theories are stopped by a drag force. The holographic description relates the confining scale in the bulk geometry with a range of physical values for the drag force in the vacuum. The vacuum drag force acting on the isolated quark directly manifests quark confinement since it prevents the quark from walking freely in the vacuum. However, analytical expressions for the drag force as a function of the quark velocity were lacking. In the present work, we propose that the vacuum drag force is given by the regularized zero-temperature limit of the corresponding thermal drag force. Within this approach, we obtain the desired analytic expressions in two different holographic models: the quadratic dilaton and the D-instanton. In both cases, we find well-behaved functions belonging to their physical range of values. \vspace{1em}
\end{abstract}
\keys{Quark-vacuum interaction, holography, drag force \vspace{-4pt}}
\pacs{   \bf{\textit{11.25.Tq, 12.38.Aw, 13.60.Hb }}    \vspace{-4pt}}
\begin{multicols}{2}

\section{Introduction}

Particle motion in a populated medium is damped due to scattering off the particles composing the medium. The averaged effect of these collisions produces the drag force acting against the motion. This situation is realized in a quark-gluon plasma where the quarks can move long distances facing other particles as obstacles. In this case, the moving quark is subjected to a drag force due to the other particles that populate the environment. Such a phenomenology can be described using gauge/gravity duality. The holographic picture was established by Gubser in \cite{Gubser:2006bz}.
Gubser’s proposal follows the AdS/CFT correspondence \cite{Maldacena:1997re,Witten:1998qj,Aharony:1999ti} that relates a conformal field theory with a higher dimensional Anti-de Sitter space-time. In the boundary gauge theory, the moving quark is dual to a trailing string in the bulk attached in the boundary. By considering this holographic picture, Gubser’s proposal was extended to the context of non-conformal AdS/QCD models at finite temperature and/or chemical potential \cite{Nakano:2006js, Talavera:2006tj, Zhang:2018rff, Xiong:2019wik, Tahery:2020tub}.

At zero temperature quarks are confined inside hadrons and no single quark can walk alone for long distances. Still, there are at least two reasons for considering the problem of a single, isolated quark walking in a vacuum. First, one can think of the situation where there is only one quark in the whole universe so that there are no other quarks to form hadrons. Second, it is observed in heavy ion collision experiments that some quarks are thrown too hard that, for some time, they lose correlation with their partners due to the
breaking of the flux tube \cite{Ali:2010tw}. The problem of a single quark walking in the vacuum of a confining gauge theory was discussed in Ref. \cite{Diles:2019jkw}. A particular feature of the AdS/QCD models is the presence of an energy scale associated with confinement at zero temperature. In Ref. \cite{Diles:2019jkw} is shown that the confinement scale allows for a drag force on the single quark moving in the vacuum and establishes an upper bound for its physically acceptable absolute value. Such a drag force at the boundary comes from the momentum flow through the trailing string stretching in the bulk. It happens that analytic expressions for the vacuum drag force as a function of quark velocity were lacking. The present work complement this holographic description of vacuum drag force by providing a step-by-step procedure to obtain the drag force as a function of quark velocity in the vacuum. To do so, we establish the vacuum drag force as the regularized zero-temperature limit of the thermal drag force. We apply our proposal in two different AdS/QCD models: the quadratic dilaton \cite{Andreev:2006ct} and the D-instanton background \cite{Liu:1999fc}.

The paper is organized as follows. In section {\bf 2}, we review the holographic calculation of the drag force in the finite temperature. In section {\bf 3}, we define the procedure to obtain the vacuum drag force by taking the regularized zero-temperature limit and show to the soft wall and D-instanton models. We conclude in Section {\bf 4}.

\section{Drag force from holography}

Here we provide a brief review of the holographic calculation of the drag force in the context of finite temperature AdS/QCD. We represent the finite temperature bulk geometry by an AdS black-brane deformed by the
presence of a dilaton:
\begin{equation}
    ds^2=\frac{R^2}{z^2}e^{\Phi(z)}\left[-f(z)dt^2 + (d\vec{x})^2 + \frac{dz^2}{f(z)}\right],
\end{equation}
with $z\in(0,z_h)$, $f(z)$ is the blackening factor satisfying $f(z_h)=0$, and the functional form of $\Phi(z)$ is provided by specifying the holographic model.

The dual picture of the moving quark is given by an open string with one endpoint attached to the quark location at the conformal boundary $(z\to 0)$. We use the string parameterization $\sigma_0=t,~\sigma_1=z$ and realize the string profile by the ansatz $X^\mu(t,z)=(t, vt+\xi(z),0,0,z)$. The information of the world-sheet deformation is encoded in the function $\xi(z)$. The boundary condition $\xi(0)=0$ is imposed to ensure that the string endpoint is attached to the quark location. Under this gauge choice, the embedding metric $\gamma_{ab}$ reads
\begin{align}
    ds^2 &= \gamma_{ab}d\sigma^a d\sigma^b =  \frac{R^2}{z^2}e^{\Phi(z)}\times  \nonumber\\
 &\bigg[-f(z)\left(1-\frac{v^2}{f(z)}\right)dt^2 + 2v\xi'(z)dtdz  \nonumber\\
 &+\frac{1}{f(z)}\left(1+f(z)\xi'2(z)^2)dz^2 \right)\bigg].
\end{align}
The Nambu-Goto action is given by the determinant of
the embedding metric, using the explicit form of $\gamma_{ab}$
we arrive at
  \begin{equation}
      I_{NG}=-\frac{1}{2\pi\alpha'}\int d^2\sigma\frac{R^2}{z^2}e^{\Phi(z)}\sqrt{1-\frac{v^2}{f}+f\xi'^2 }.
  \end{equation}
The drag force on the boundary quark is given by the flow of the $x$ component of the linear momentum along the holographic $z$ direction:
   \begin{equation}
F_{drag} = \Pi^z_{~x} = \frac{\delta I_{NG}}{\delta (\partial_z x)}.       
   \end{equation}
   In our case at hand,
\begin{equation}
   \Pi^z_{~x} = -\frac{R^2}{2\pi\alpha'}\frac{e^{-\Phi(z)}}{z^2}\frac{f\xi'}{\sqrt{1-\frac{v^2}{f}+f\xi'^2 }}.  
\end{equation}
The on-shell conservation of $\Pi^z_{~x}$ leads to a first order equation for $\xi$:
\begin{equation}\label{eom}
    \frac{d\xi}{dz} =\frac{2\pi\alpha'}{f}\sqrt{\frac{f-v^2}{\frac{R^4e^{-2\Phi}}{z^4} -(2\pi\alpha' \Pi^z_{~x})^2 }}.
\end{equation}
Requiring $\xi$ to be real-valued for $z\in(0,z_h)$ means that the denominator inside the square root should change the sign with the numerator. We call the turning point $z_c$ , defined by $f(z_c)=v^2$. The conserved momentum is obtained
\begin{equation}
  F_{drag} = \Pi^z_{~x} = - \frac{e^{\Phi(z_c)}}{2\pi\alpha'z_c^2}\sqrt{f(z_c)}.  \label{dragforceT}
\end{equation}

It is important to remark that the presence of an event horizon is considered from the beginning of the discussion, leading to the above equation. The analogous discussion starting from a bulk geometry with
no event horizon does not provide an expression for the drag force. Instead, one finds a finite range for the allowed values that drag force can assume. In the following, we will start from the finite temperature drag force to establish analytic expressions for the vacuum drag force.

\section{The zero temperature limit}

The reality condition imposed on equation (\ref{eom}) leads to an analytic expression for the drag force at finite temperature. The temperature information is encoded into the critical value $z_c$ where the $\frac{0}{0}$ point is crossed. We remind that the blackening function also depends on the horizon location, and then
\begin{equation}
    f(z;z_h)=v^2\rightarrow z_c(z_h)=f^{-1}(v^2;z_h).
\end{equation}

The precise information of the temperature comes from $T(z_h)=\frac{1}{|f'(z_h)|}$. For the cases of interest, it will hold that $T\to 0$ is equivalent to $z_h\to\infty$. With these general results at hand, we will define the analytic expressions for the drag force in the vacuum as the regular part of the zero-temperature limit, or equivalently the $z_h\to\infty$ limit of equation (\ref{dragforceT}). In this sense, the drag force due to the vacuum is the zero temperature remnant from the thermal drag force. A similar approach known as opacity expansion is used to discuss quark fragmentation into jets \cite{Armesto:2003jh, Zhang:2007ai}.

\subsection{AdS deformed by quadratic dilaton}

The AdS space deformed by a quadratic dilaton is shown to be confining \cite{Andreev:2006ct}, and according to \cite{Diles:2019jkw} a single quark on its vacuum is subjected to a drag
force. Here we apply the proposal for defining the vacuum drag force as the regularized zero-temperature limit of the thermal drag force. In the present bottom-up AdS/QCD model, the bulk geometry is an AdS-Schwarzchild black brane deformed by the quadratic dilaton $\phi(z)=k^2z^2$. The blackening factor is $f(z) = 1-(\frac{z}{z_h})^4$, leading to
\begin{equation}
    z_c^2=z_h^2\sqrt{1-v^2},~~T=\frac{1}{\pi z_h}.
\end{equation}

Inserting the above expressions in eq.(\ref{dragforceT}) one obtain the drag force as a function of the temperature:
  \begin{equation}
      F_{drag}(v,~T) =  \frac{\pi^2 \,T^2 \,R^2}{2\,\alpha'}\,e^{\frac{k^2\sqrt{1-v^2}}{2\pi^2 T^2}}\, \frac{v}{\sqrt{1-v^2}}.
 \end{equation}
 To identify the regular part when $T\to 0$ we expand  $T^2e^{\frac{\alpha}{T^2}}\sim T^2+\alpha+O(\frac{1}{T^2})$. The divergent terms correspond to the $O(\frac{1}{T^2})$ that is just ruled out. Then we take the zero-temperature limit of the remaining terms and obtain
\begin{equation}
    F_{vac}(v) = -\frac{R^2k^2}{4\pi^2\alpha'}v.
\end{equation}
 Note that as $v \leq 1$, the expression above respect the allowed range predicted in  \cite{Diles:2019jkw}. However, this drag
force is not linear in the relativistic momentum, and as a consequence, the momentum damping will not be exponential.

\subsection{D-instanton background}

An interesting AdS/QCD scenario is given by the D-instanton background \cite{Zhang:2018rff, Gwak:2012ht}. This scenario is known
to be confining in the sense of a linear $q\bar{q}$ potential. In \cite{Zhang:2018rff}, the finite-temperature geometry is explored and
the drag force acting on the single quark moving in the dual hot medium is given by
\begin{align}
     F_{drag}(v) =& -\sqrt{1+\frac{q}{\pi^4 R^8 T^4 }\ln \left(\frac{1}{v^2}\right)}\nonumber \\
     &\frac{\pi T^2\sqrt{\lambda}}{2}\frac{v}{\sqrt{1-v^2}}.
\end{align}
 
 This time, the zero-temperature limit is obtained without any regularization, leading to
 \begin{equation}
    F_{vac}(v) = -\frac{\sqrt{ q\ln(v^{-2})}}{2\pi\alpha' R^2} \frac{v}{\sqrt{1-v^2}}.
\end{equation}

The obtained drag force in the vacuum is bounded from above when $v\to1$. One can easily check that the above expression respects the upper bound predicted in \cite{Diles:2019jkw} for the vacuum drag force by taking the zero-temperature limit for the D-instantom background geometry \cite{Gwak:2012ht}. This result represents one more example where we can obtain an analytical expression for the drag force in the vacuum of a confining gauge theory with holographic dual.

\section{Conclusions}

In this work we have updated the proposal of Ref. \cite{Diles:2019jkw} by providing a step-by-step procedure to obtain the vacuum drag force as a function of quark velocity. To show how the procedure works we considered two different AdS/QCD models: the AdS deformed by a quadratic dilaton and the D-instanton background. For the Anti-de Sitter space deformed by the quadratic dilaton we face a divergent zero-temperature limit and regularize the expression by subtracting positive powers of $\frac{1}{T}$. This proposal is analogous to the treatment given in Ref. \cite{Tong:2009np} for the Casimir force. On the other hand, in the D-instanton background, we found a convergent zero temperature limit for the thermal drag force. The expression obtained for the vacuum drag force in the D-instanton background is found to be continuous in the range $v\in[0,1]$ and respect the upper bound imposed by the zero temperature geometry.

 The results presented paves one more step in the discussion started in \cite{Diles:2019jkw} concerning the role of confinement in the path of a single quark as well as the viscous nature of a confining vacuum from a holographic perspective. We also remark that our proposal for obtaining vacuum drag force can, in principle, be applied in other confining holographic models of QCD.

\end{multicols}

\medline

\begin{multicols}{2}

\end{multicols}

\end{document}